\documentclass[pre,superscriptaddress,twocolumn,showpacs,showkeys]{revtex4}
\usepackage{epsf,amsmath,amssymb,wasysym,color}
\usepackage[pdftex]{graphicx}
\newcommand{\be}{\begin{equation}}
\newcommand{\ee}{\end{equation}}

\begin{document}
\title{Sandpiles on Multiplex Networks}
\author{Kyu-Min Lee} 
\author{K.-I. Goh}
\email{kgoh@korea.ac.kr}
\author{I.-M. Kim} 
\affiliation{Department of Physics and Institute of Basic Science, Korea University, Seoul 136-713, Korea}

\date{\today}

\begin{abstract}
We introduce the sandpile model on multiplex networks with more than one type of edge and investigate its scaling and dynamical behaviors. We find that the introduction of multiplexity does not alter the scaling behavior of avalanche dynamics; the system is critical with an asymptotic power-law avalanche size distribution with an exponent $\tau = 3/2$ on duplex random networks. The detailed cascade dynamics, however, is affected by the multiplex coupling. For example, higher-degree nodes such as hubs in scale-free networks fail more often in the multiplex dynamics than in the simplex network counterpart in which different types of edges are simply aggregated. Our results suggest that multiplex modeling would be necessary in order to gain a better understanding of cascading failure phenomena of real-world multiplex complex systems, such as the global economic crisis.
\end{abstract}
\maketitle

\section{Introduction}
Complex network theory has provided a useful tool for studying the structure and the dynamics of complex systems \cite{newman,havlin}. Various dynamical processes occuring on networks have also been extensively studied \cite{dynamics}. An important class of dynamical processes is the so-called cascading failure dynamics, addressing how the failure of nodes induces other nodes' failure over the network. Various real-world phenomena such as blackouts of electrical power grids \cite{motter02, sachtjen}, Internet congestion cascades \cite{motter02,ejlee}, or global economic crises \cite{kyumin} can be modeled under the framework of cascading failure on networks. Prototypical models of cascading failure include the sandpile model \cite{btw}, the threshold cascade model of behavioral adoptions \cite{watts02}, and the capacity-based overload cascade model \cite{motter02}.

Most studies thus far have focused on networks with a single type of link, the so-called simplex networks. In many real-world complex systems, nodes participate in more than one type of interaction with other nodes in the system; People in society have friendship, family ties, professional collaborative relations, {\it etc}. Countries in the global economy interact through various financial and political channels; they are multiplex networks. Different interaction channels do not always operate in the same way, nor do they operate in a completely autonomous way. Therefore, the multiplexity of networks introduces another nontrivial structural factor in complex networks, the dynamic consequence of which is to be understood. 
In this perspective, there have been recent studies related to the multiplexity of networks in the form of interdependency between network layers \cite{buldyrev10}, interacting networks \cite{leicht}, and multi-relational structure of online social networks \cite{szell}. These studies showed that the multiplexity presents nontrivial structural patterns and connectivity \cite{leicht,szell,kyumin11} and can drastically change the robustness properties of interdependent network systems \cite{buldyrev10}. This suggests that the simplex network framework would generically be incomplete and thus one needs to consider the multiplex network framework for a better understanding of complex systems. 

Each type of interaction in multiplex networks defines a network layer; for example, a social network may comprise a friendship layer, a family tie layer, {\it etc.} A change of a node's state may be determined by intralayer dynamics; for example a person in a social network may adopt a behavior (using a smartphone app, say) by the peer pressure from his/her friends or by a demand due to the work environment alone. However, the potential influence of his/her state change is not confined within that layer; even though you adopted the app due to your work demand, your usage of the app may spread to your friends or family as you may recommend it through your friendship and family tie layers. In this way, the layers of a network get coupled and become interdependent. Such a multiplex coupling is generically not equal to a simple aggregation of link types, thereby leading to nontrivial dynamical consequences. 

In this paper, we apply multiplex modeling to a simple model of cascading failure, the sandiple model, and investigate its dynamical impact.
We begin by reviewing the sandpile dynamics on random, simplex networks in Sec.~II. The multiplex sandpile model is introduced in Sec.~III, and the principal results will be presented in Sec.~IV. Other model variants are considered in Sec.~V, and the paper concludes in Sec.~VI.

\section{Sandpiles on random networks}

The sandpile model was originally introduced by Bak, Tang, and Wiesenfeld (BTW) in 1987 \cite{btw}. Since then, this model has been an archetypical theoretical model for investigating cascading failure dynamics and self-organized criticality \cite{soc}. The dynamics proceeds as follows. i) Each node is assigned a prescribed threshold height, usually taken to be its coordination number $\{c_x=z_x\}$. ii) At each step, a grain is added at a randomly chosen node, say $x$, raising the node's height by one, $h_x\leftarrow h_x+1$. iii) If the height of the node becomes equal to or exceeds its threshold, that is $h_x\ge c_x$, the node becomes unstable and topples $c_x$ grains to its neighbors, such that $h_x\leftarrow h_x-c_x$ and $h_y\leftarrow h_y+1$ for $y$ being the neighbor of $x$. iv) If this toppling event induces instability in some neighbors, they also topple in the same way; a cascade of toppling events (an avalanche) occurs until there remain no more unstable nodes in the system. v) Repeat ii)--iv).
BTW showed that without a fine-tuning of the external parameters, this model could exhibit a scale invariance in the form of a power-law distribution of avalanche sizes (the number of nodes participating in a given avalanche), 
\be P(s)\sim s^{-\tau}, \ee hence, the name self-organized criticality \cite{btw}. 

In Euclidean lattices in low dimensions, the BTW sandpile model exhibits nontrivial critical behaviors \cite{soc}. Its dynamics on random networks, however, can be understood more easily, following the mean-field behaviors, $P(s)\sim s^{-3/2}$ \cite{alstrom,bonabeau}. Later, the model has been considered on the scale-free networks with a power-law degree distribution, $P(k)\sim k^{-\gamma}$, where the degree $k$ is the number of links of a node \cite{goh03,jkps04,physa04,physa05}. For the BTW sandpile model with $c_i=k_i$ on scale-free networks, the unconventional mean-field behavior was obtained such that the avalanche size exponent $\tau$ is given by \cite{goh03}
\be \tau=\left\{\begin{array}{ll}3/2 & (\gamma>3), \\ \gamma/(\gamma-1) \quad & (2<\gamma<3).\end{array}\right. \ee
Here, we briefly present the analytical approach based on the branching process approximation leading to these results \cite{goh03,jkps04}.
Assuming that the avalanche proceeds without forming a loop, it can be mapped onto a branching process. In this mapping, the probability $q_k$ to make a $k$-branch $(k\ge1)$ at each branching event is given by
\be q_k = \frac{kP(k)}{\langle k\rangle} \frac{1}{k}~. \ee
The first factor comes from the likelihood that a node with degree $k$
would gain a grain from its unstable neighbor, which is proportional to
its degree. The second factor comes from the assumption that the timing
at which the node gets a grain from its neighbor is random, such that
the probability to become unstable (that is, $h=k-1$ at the moment) is $1/k$. $q_0$ is given by the normalization as $q_0=1-\sum_{k=1}^{\infty}q_k$, and the mean branching number $B\equiv\sum_{k=0}^{\infty}kq_k$ is obtained as 
\be B = \sum_{k=0}^{\infty}kq_k=\sum_{k=1}^{\infty}\frac{kP(k)}{\langle k\rangle}=1~.\ee
Therefore, the branching process is critical.
The asymptotic (large $s$) behavior of the avalanche size distribution  $P(s)$
can be obtained from the singular behavior near unity in the corresponding generating function \cite{jkps04}. 
On scale-free networks, the branching probability decays algebraically, $q_k \sim P(k)\sim k^{-\gamma}$, which gives rise to a singular term $\sim(1-\omega)^{\gamma-1}$ in its generating function, leading to the asympototic power-law behavior of {\bf$P(s)$} for $\gamma<3$ as given by Eq.~(2). Thus, the behavior of the branching probability $q_k$ provides an essential clue to the sandpile dynamics. 

\section{Sandpile model on multiplex networks}

\begin{figure}[b]
\centering
\includegraphics[scale=1.,width=1.\columnwidth]{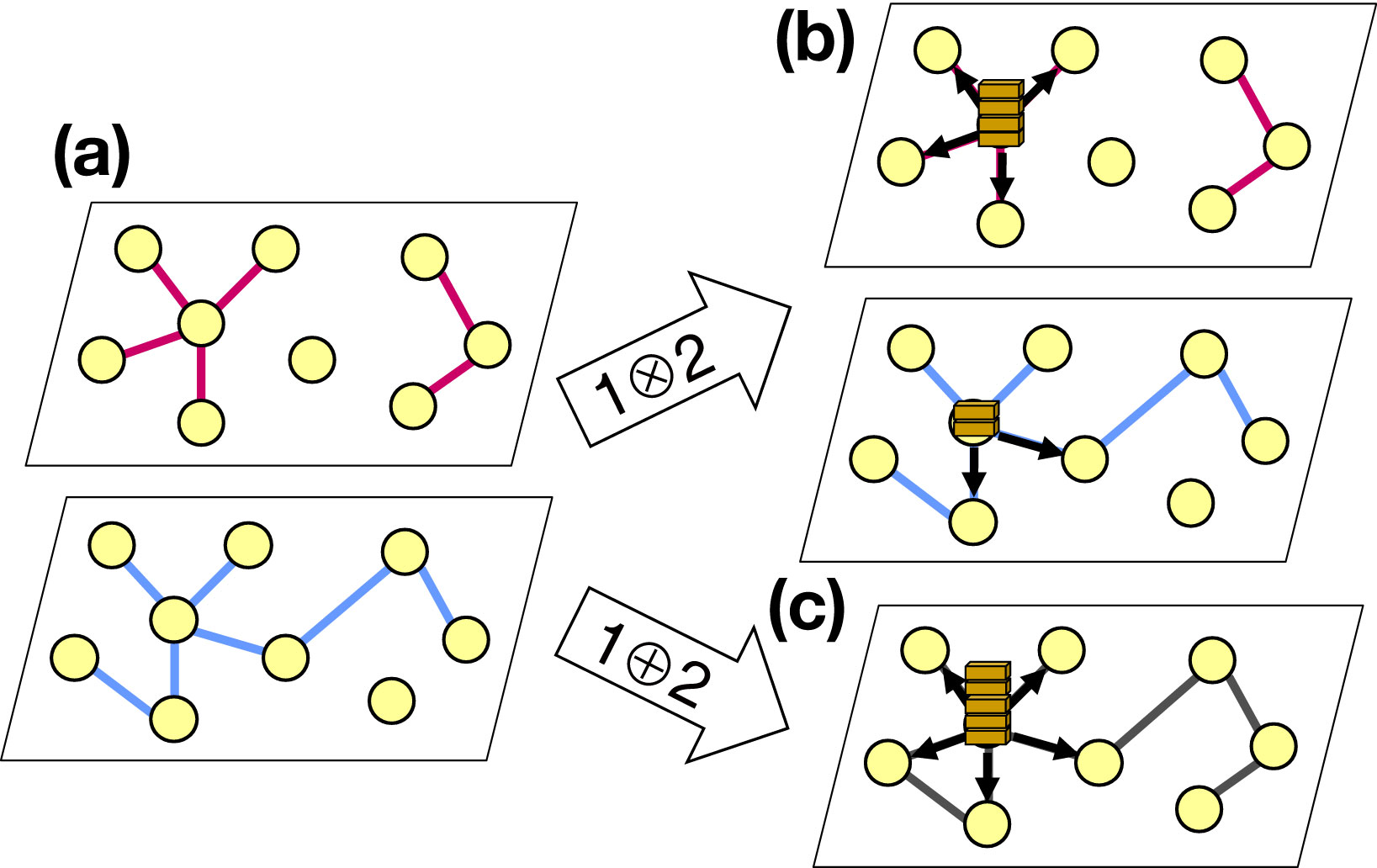}
\caption{(Color) Schematic illustrations of the model dynamics. (a) A network is composed of two layers of red (top) and blue (bottom) interactions, respectively. (b) In multiplex dynamics, the individuality of different interactions is respected; thus, each node has two piles of grains that follow the multiplex 1$\otimes$2 dynamics. (c) In the simplex 1$\oplus$2 case, two types of interactions are aggregated, and the conventional network sandpile is run on the aggregated network. }
\end{figure}

In this paper we generalize the BTW model on multiplex networks. 
In multiplex networks, each type of interaction functions as a network layer within which the dynamics takes place as if in the usual single network case. However, the state change of a node in one layer applies also to other layers, giving rise to couplings between layers' dynamics. Such an interlayer coupling induces interdependency between layers and renders the multiplex dynamics rather nontrivial \cite{buldyrev10,bjkim,brummitt}. To illustrate the effect of multiplexity in its simplest setting, we consider in the paper multiplex networks with two kinds of interactions or a {\em duplex} networks.

Specifically, the BTW sandpile model on duplex networks is run as follows: 
i) Each node $x$ is assigned the thresholds $c_{\ell,x}$ equal to its degree for each layer; that is, $c_{\ell,x}=k_{\ell,x}$ for $\ell=1$ and $2$. ii) At each step, a grain is added to a randomly chosen node $x$ in a randomly chosen layer $\ell$; $h_{\ell,x} \leftarrow h_{\ell,x} + 1$. iii) If the height of this node reaches or exceeds its threshold in {\em any} layer, that is, either $h_{1,x} \geq c_{1,x}$ or $h_{2,x} \geq c_{2,x}$, the node becomes unstable, initiating an avalanche. The layer in which the threshold is exceeded is called the unstable layer. iv) Each unstable node topples on {both} layers; in the unstable layers, it topples $c_{\ell,x}$ grains, one for each neighbor, whereas it topples only $h_{\ell',x}$ grains in the other layer, giving one grain to each of a random subset of $h_{\ell',x}$ neighbors [Fig.~1(b)]. That is, 
\be h_{\ell,x}\leftarrow h_{\ell,x} - \min[c_{\ell,x},h_{\ell,x}]\quad\quad\textrm{(1$\otimes$2 model)}, \ee 
and the height of all the nodes receiving a grain increases by one. v) If this toppling event causes other nodes to be unstable, these unstable nodes topple in the same way as in iv). vi) Repeat iv)--v) until there are no more unstable nodes in the system, and the avalanche ends. During each toppling, the grain is lost with a small probabilty, taken to be $f = 10^{-4}$ in this paper, which is necessary to prevent overloading the system. 
We call this multiplex dynamics model the 1$\otimes$2 model.

In our 1$\otimes$2 multiplex setting, a node's failure (toppling) due to a specific layer dynamics induces its simultaneous failure in other layers regardless of its conditions in those layers. This setting is motivated by a number of real-world situations such as the global economy, in which one country's recession due to an imbalance in the trade system can affect the country's failure in a credit network, for example. The 1$\otimes$2 model is a stylized model of such scenarios, aiming to address the effect of multiplexity. To highlight the effect of multiplexity in the 1$\otimes$2 model, we compare it to a non-multiplex, or simplex, counterpart of the model, which we call the 1$\oplus$2 model. In the 1$\oplus$2 model, we ignore the individuality of different interactions and simply overlay two layers [Fig.~1(c)]
to form a simplex aggregate network, on which the conventional BTW model is run. In the next section, we study and compare the avalanche dynamics of these two models. 

\section{Results}

The primary quantity of interest in sandpile dynamics is the avalanche size distribution $P(s)$, giving the likelihood that an avalanche of size $s$ is to occur in the steady state of the model \cite{btw,soc}. We meaure the avalanche size in terms of the number of nodes becoming unstable during an avalanche. A related quantity is the number of toppling events during an avalanche, which scales similarly when the formation of loops during an avalanche is negligible, which is the case for sandpile dynamics on random graphs \cite{goh03}. One can also measure the distribution of avalanche durations, which also follows a power law, whose exponent is related to the avalanche size exponent by a scaling relation \cite{soc}. In this work we focus exclusively on the avalanche size distribution.

\begin{figure}
\centering
\includegraphics[width=\columnwidth]{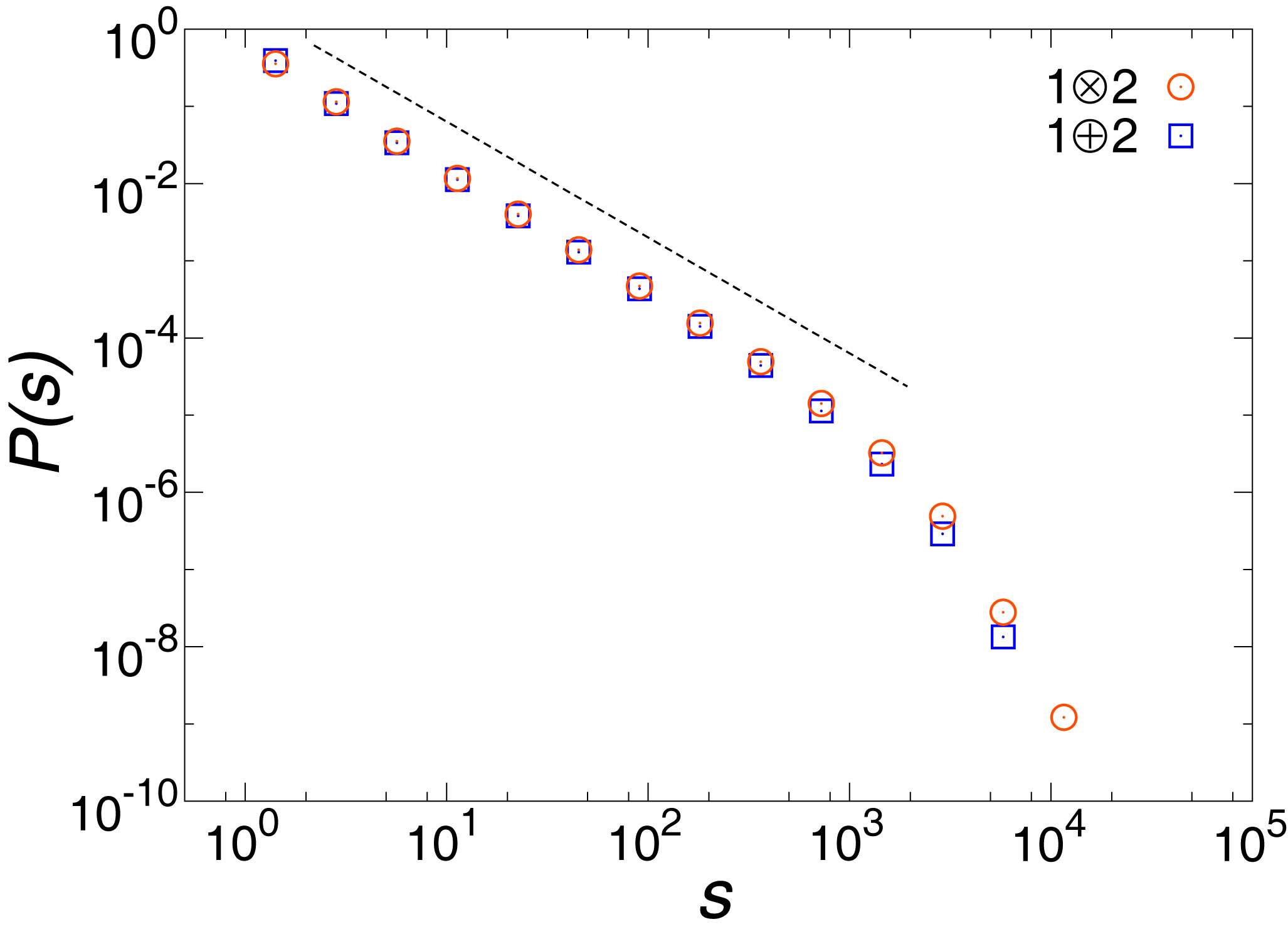}
\caption{(Color online) Avalanche size distribution $P(s)$ of the 1$\otimes$2 (circle) and the 1$\oplus$2 (square) model on duplex ER networks with $N=10^6$ and mean degree $z=4$ per layer. The slope of the dotted line is $-3/2$ and is drawn for the eye. Data are logarithmically binned.}
\end{figure}

\subsection{On Duplex Erd\H{o}s-R\'enyi Layers} 
We first consider the sandpile dynamics on multiplex networks with two Erd\H{o}s-R\'enyi (ER) \cite{er} layers of $N=10^6$ nodes and the equal mean degree $z=4$. Each node has degrees $k_1$ and $k_2$ in the first and the second layers, respectively, each of which is a Poisson random variable with the mean equal to $z=4$. To obtain the avalanche size distribution and related statistics, we sampled $10^6$ successive avalanches after the initial transient period.
For the simplex 1$\oplus$2 model, the substrate network is nothing but an ER network with mean degree $\langle k\rangle=2z=8$. Therefore, the 1$\oplus$2 model is trivially critical, with the standard mean-field exponent $\tau_{1\oplus2}=3/2$ \cite{bonabeau,alstrom}, which is also confirmed by numerical simulations (Fig.~2). 

For the multiplex 1$\otimes$2 model, the sandpile dynamics would become more complicated due to layers' coupling or interdependency. Even the criticality of the model is not obvious {\it a priori}. The numerical simulation of the 1$\otimes$2 model, however, showed that its avalanche size distribution is essentially identical to that of the 1$\oplus$2 model; $P(s)$ follows a power law, that is, the model is still critical, with the same mean-field exponent $\tau_{1\otimes2}=3/2$ (Fig.~2).  Such a robust mean-field behavior was also reported recently for sandpile models on interdependent networks in somewhat different settings \cite{charlie}. 

\begin{figure}[t]
\centering
\includegraphics[width=\columnwidth]{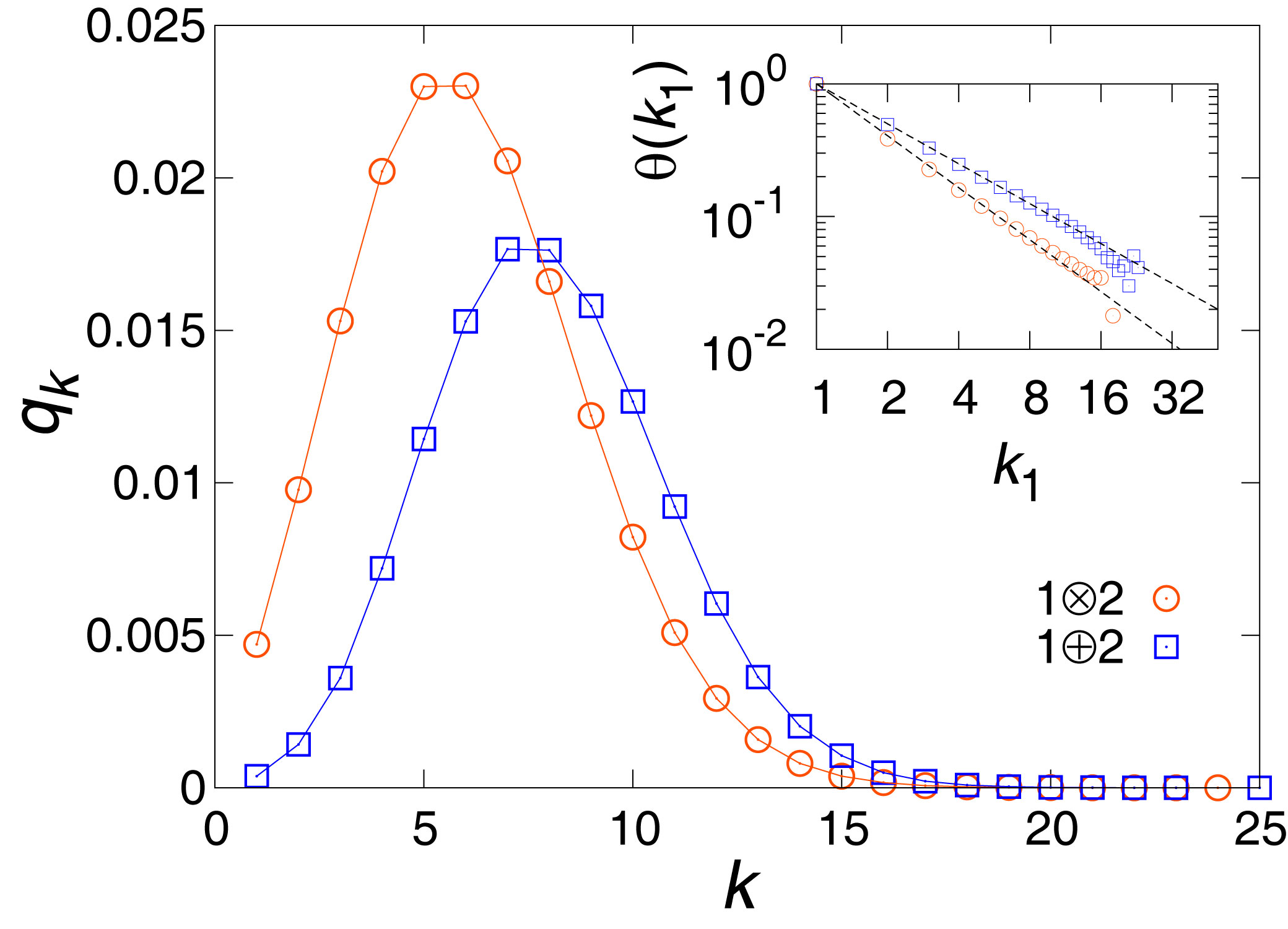}
\caption{(Color online) Branching probability $q_k$ for the 1$\otimes$2 (circle) and the 1$\oplus$2 (square) model. (Inset) Plots of the factor $\theta(k_1)$ in Eq.~(7) for the two models. The slopes of the dotted lines are $-1.0$ (upper) and $-1.3$ (lower), respectively and are drawn for the eye.}
\end{figure}

Although the two models show similar $P(s)$ with the same exponent $\tau$, the specific sandpile dynamics underlying this distribution are different, given the presence/absence of multiplexity. Compared to the simplex case that can be understood in terms of the framework presented in Sec.~II, the multiplex dynamics follows a more complicated cascade process. To provide more detailed insight, we again employ the branching process approximation following Sec.~II.
For the simplex 1$\oplus$2 case, the branching probability is given by, from Eq.~(3), 
\be q_k^{1\oplus2} = \frac{P_{1\oplus2}(k)}{\langle k\rangle}~, \ee
where $P_{1\oplus2}(k)$ is the degree distribution of the aggregate network of two layers. Equation.~(6) is confirmed by numerical simulations [Fig.~3(a)]. For the multiplex 1$\otimes$2 model, the associated branching process can be divided into two parts: one through the unstable layer (denoted layer 1) and the other through layer 2. The latter is the consequence of multiplexity and the layers' interdependency. To obtain the branching probability into $k$ branches, one has to consider the composite branching of the $k_x$-branch in layer 1 and the $h_x$-branch in layer 2. 
Therefore, one can write 
\begin{align}
q_k^{1\otimes2} &= \sum_{k_1=1}^{k}q_{k_1}^{(1)}q_{k-k_1}^{(2)} \nonumber \\ &=\sum_{k_{1}\leq k} \frac{k_{1}P(k_1)}{\langle k_1 \rangle}\theta (k_1) \sum_{k_2 \geq k-k_1}\frac{1}{k_2}P(k_2)~.
\end{align}
The first summand gives the probability that a node with degree $k_1$ receives a grain and becomes unstable, making a $k_1$-branch. The second summation gives the probability that the node has $k-k_1$ grains in layer 2 at the moment (thereby $k-k_1$ branching), assuming again the random timing of topplings. Summing the combined factor over $k_1$, we have $q_k$. Multiplexity is clearly manifested in the second factor, but it does play a role in the first factor, too. 
The unspecified factor $\theta(k_1)$ addresses the probability that a node becomes unstable as it receives a grain, which was $1/k$ for simplex sandpile, Eq.~(3). The assumption behind it was that all heights are equally likely when encountering a toppling, as the grains are successively piled. In the multiplex model, however, between two successive receipts of grains, a node can topple due to an instability in the other layer. Accordingly, the likelihood that a node becomes unstable as it receives a grain becomes smaller than $1/k$, and we numerically find that it scales roughly as $\theta(k_1)\sim 1/k_1^{\alpha}$ with $\alpha\approx1.3$ in our simulation setting [Fig.~3(a), inset]. Therefore, the multiplex coupling induces truly nontrivial dynamics that cannot be viewed as a simple sum of two individual layers' dynamics.

The overall $q_k^{1\otimes2}$ is shown in Fig.~3(a) along with the simplex counterpart, showing that in multiplex sandpile dynamics, branching into smaller number of branches (toppling a smaller number of grains) occurs more frequently than in simplex sandpiles. Meanwhile, the mean branching number $B$ is $1$ in both cases; numerically, we measured $B^{1\oplus2}\approx1.002$ and $B^{1\otimes2}\approx1.005$. This supports the robustness of self-organized criticality. Although $q_k^{1\otimes2}$ takes a nontrivial form, its large-$k$ behavior is still dominated by the Poisson tail, which provides an explanation for the robustness of the avalanche size exponent.  

\begin{figure}[t]
\centering
\includegraphics[width=.95\columnwidth]{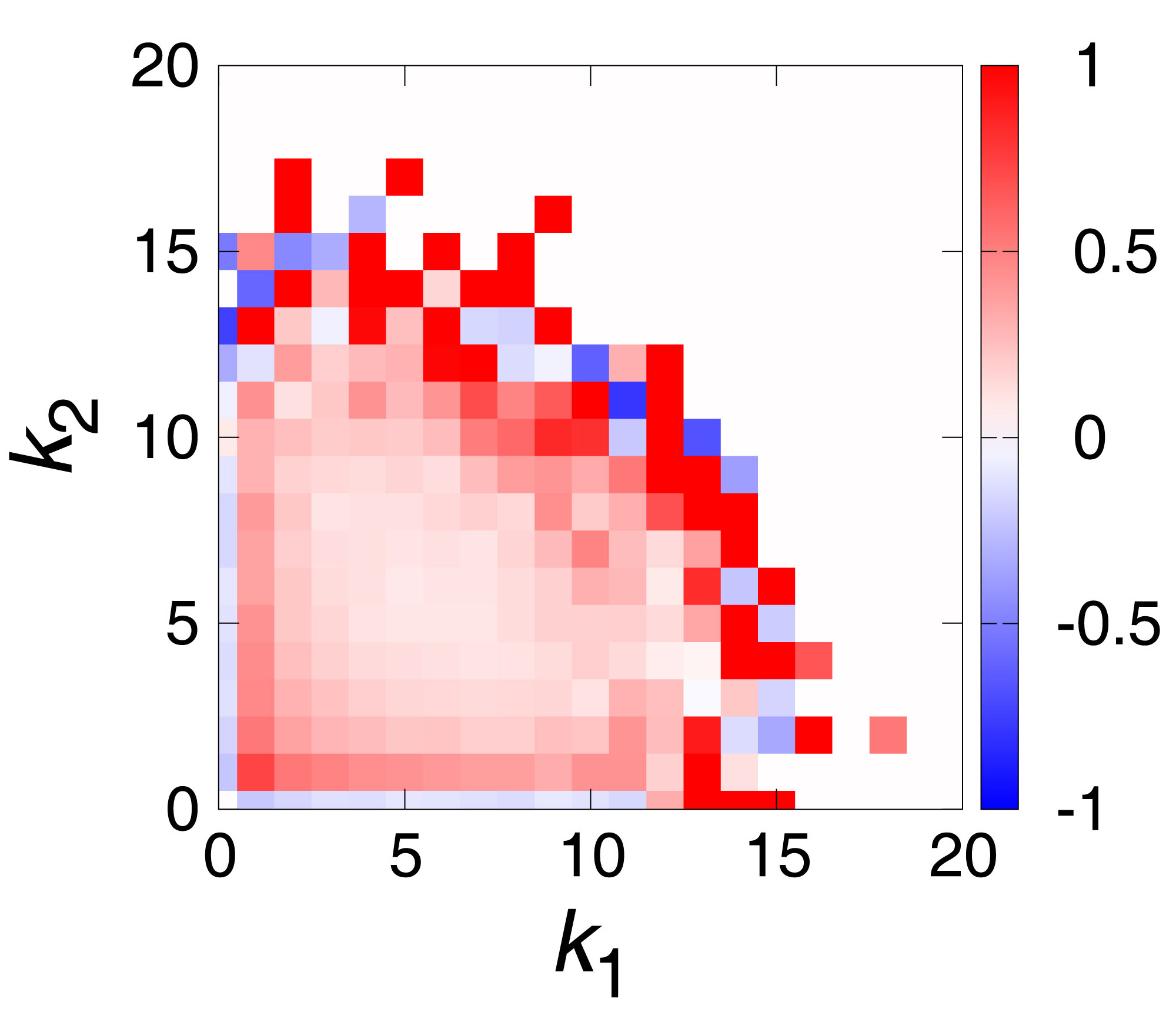}
\caption{(Color) Plot of the normalized difference between the frequency of node failure as a function of node degrees, $\phi(k_1,k_2)$, defined in the text. A positive (red) value of $\phi$ means that the node fails more frequently in the 1$\otimes$2 model than in the 1$\oplus$2 model. A negative (blue) $\phi$ indicates the opposite.}
\end{figure}

To differentiate the cascade dynamics of the two models in more detail, we measured the frequency of topplings in nodes with $k_1$ and $k_2$, $f(k_1, k_2)$ and plotted the normalized difference $\phi(k_1,k_2)=[f^{1\otimes2}(k_1,k_2)-f^{1\oplus2}(k_1,k_2)]/f^{1\oplus2}(k_1,k_2)$ in Fig.~4. One can see that most nodes fail more frequently in the 1$\otimes$2 model than in the 1$\oplus$2 model whereas the nodes with links in only one layer (either $k_1=0$ or $k_2=0$) fail more in the simplex dynamics. This result suggests higher vulnerability of most nodes in multiplex sandpile cascades than in simplex scenarios. This picture is in line with recent studies indicating higher vulnerability of interdependent networks \cite{buldyrev10} and multiplex cascade models \cite{charlie}.

\begin{figure}[t]
\centering
\includegraphics[width=1.\columnwidth]{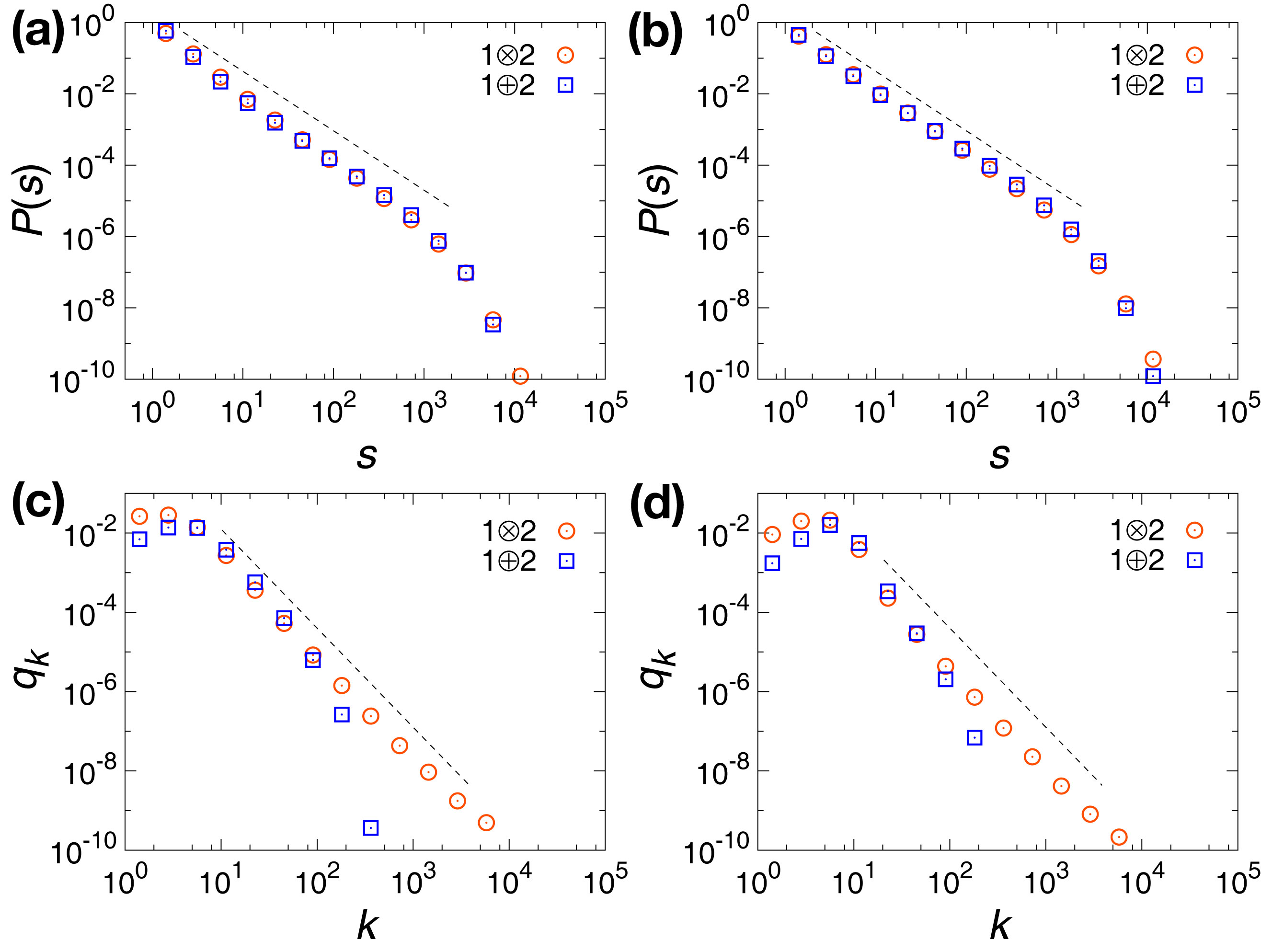}
\caption{(Color online) (a, b) Avalanche size distribution $P(s)$ of the sandpiles on duplex networks with (a) two SF layers and (b) one SF and one ER layers. The degree exponent of the SF layers is $\gamma=2.5$. The slopes of both dotted lines are $-5/3$ and are drawn for the eye. (c, d) Branching probability $q_k$ on (c) SF-SF and (d) SF-ER duplex networks. The slopes of both dotted lines are $-2.5$ and are drawn for the eye.}
\end{figure}

\subsection{On Scale-free Layers}
We have also performed numerical simulations on multiplex networks with i) two scale-free (SF) layers and ii) one ER and one SF layer. For the SF layer, we used the static model \cite{goh01} with the same mean degree $z=4$ and the degree exponent $\gamma=2.5$. For the simplex dynamics, the degree distribution of aggregate networks in both cases has a power-law tail with exponent $\gamma=2.5$, which leads to the predicted avalanche size exponent $\tau=5/3$ according to Eq.~(2). The multiplex dynamics is still critical ($B^{1\otimes2}\approx1.03$ and $B^{1\oplus2}\approx1.005$), having the same power-law avalanche size distribution in both cases [Figs.~5(a,b)]. The detailed cascade dynamics are different; the branching probability $q_k$ is larger for very small and very large $k$, but smaller for intermediate $k$ in the multiplex dynamics than in simplex ones [Figs.~5(c,d)]. This result suggests an interesting picture that hubs becomes more vulnerable in multiplex sandpiles. In an ordinary (simplex) sandpile, hubs play the role of a shock-absorbers, by tolerating a large amount of stress (grains) without failing; in multiplex dynamics, hubs fail more often as there is an additional chance of failure due to its interdependency with a low degree node. Therefore, although the multiplex and the simplex dynamics do not differ in terms of scaling behaviors, the microscopic picture of cascade dynamics at the individual node level does show differences that may have implications for risk assessment and aversion strategy for cascading failure in real-world phenomena.

\section{Modifications of the model}

\begin{figure}[t]
\centering
\includegraphics[width=1.\columnwidth]{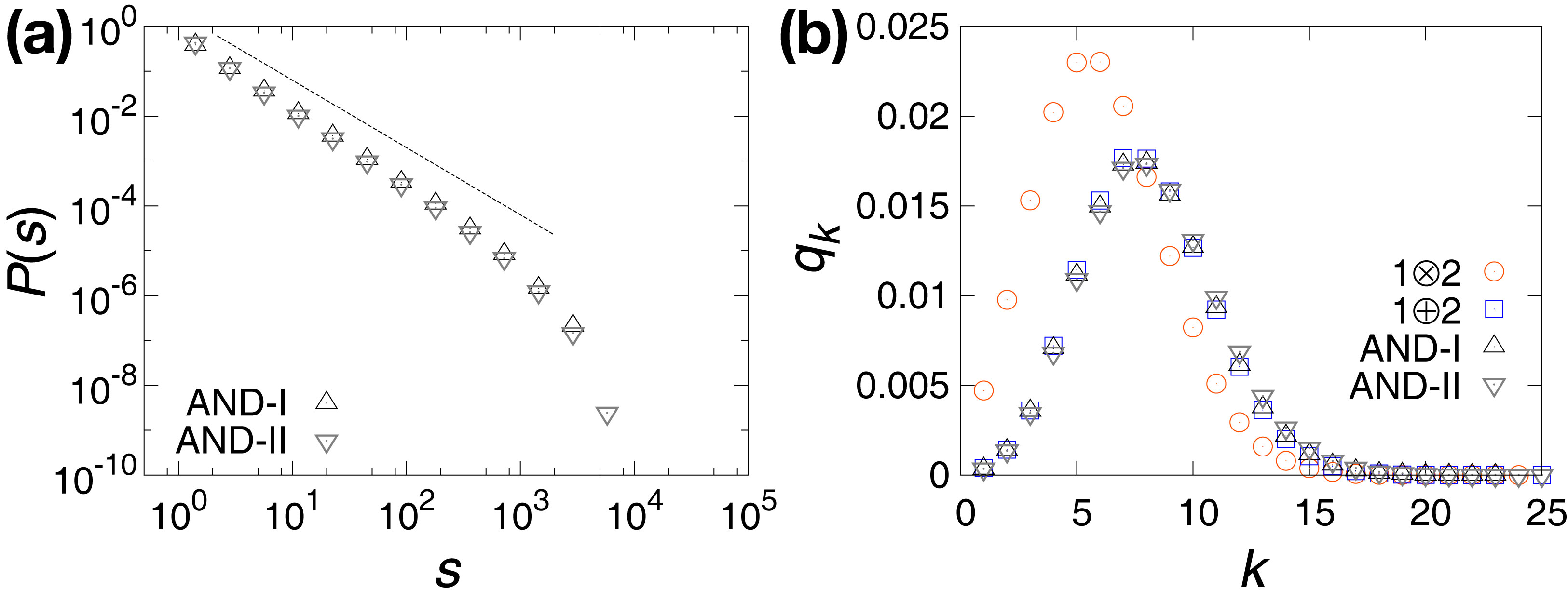}
\caption{(Color online) (a) Avalanche size distribution $P(s)$ of the AND-I (triangles) and AND-II (down triangles) models. The slope of the dotted line is $-3/2$ and is drawn for the eye. (b) Branching probability $q_k$ for the two models, compared with those of the 1$\otimes$2 (circles) and the 1$\oplus$2 (squares) models. }
\end{figure}

In this section, we consider other variants of the BTW sandpile model on multiplex networks. The first variant is the one with a different rule iii) of the 1$\otimes$2 model in Sec.~III. In the new model, a node fails (topples) if it is unstable in {\em both} layers, that is, $h_{\ell,x}\ge c_{\ell,x}$ for both $\ell=1, 2$. When a node fails, it topples $c_{\ell,x}=k_{\ell,x}$ grains in each layer. We call this version the AND-I model. We also consider another rule such that a node topples all its grains. We call this variant the AND-II model. Therefore, for these variants we have instead of Eq.~(5), 
\begin{align}
h_{\ell,x} & \leftarrow h_{\ell,x}-c_{\ell,x} &\textrm{(AND-I model),}\nonumber\\
h_{\ell,x} & \leftarrow 0 & \textrm{(AND-II model)}.
\end{align}
For the AND-II model, when a node topples more than $k_{\ell}$ grains in the layer $\ell$, it first distributes $k_{\ell}$ grains to its neighbors and the remaining $h_{\ell}-k_{\ell}$ grains to randomly chosen neighbors.
The scaling behavior is found to be still robust with respect to both variants of multiplex sandpile dynamics. They are critical ($B^{\textrm{AND-I}}\approx 1.003$ and $B^{\textrm{AND-II}}\approx1.06$) and have the same avalanche size exponent $\tau=3/2$ on duplex ER networks [Fig.~6(a)]. Interestingly, the branching probability $q_k$ for these two variants is found to be the same as that of the simplex 1$\oplus$2 model [Fig.~6(b)].

The final variant we consider is to change rule iv) of the 1$\otimes$2 model such that a node fails if it is unstable in any layer, but the unstable node topples $c_{\ell,x}$ grains in both layers even when its current height is less than the threshold in layer 2. That is, we have, 
\begin{align} 
h_{\ell,x} & \leftarrow \max[h_{\ell,x}-c_{\ell,x},0] \quad\quad \textrm{(1$\circledast$2 model)}
\end{align}
for both $\ell$, and the height of all the neighbors of $x$ increases by one. We symbolized this as the 1$\circledast$2 model, to distinguish it from the original multiplex 1$\otimes$2 model.
In this version, grains are not conserved, but produced, during the avalanche. This setting is less physical, but can model situations in which the failure due to interdependency induces the full stress ($c_{2,x}$) rather than the current stress level. We found that in this model, the system quickly enters into an infinite avalanche, in which the number of grains increases abruptly and eventually the avalanche covers the whole system, after which a finite fraction of nodes topple {\em ad infinitum} [Figs.~7(a-c)]. The branching analysis indeed shows that the mean branching number becomes larger than one, increasing up to $B\approx4$ [Fig.~7(d)]. Therefore, the branching process becomes highly supercritical and can continue forever. The excess stress (grains) generated by interdependency regenerates further excess stress, and the system falls into the trap of an infinite sequence of cascading failure. The system is no longer critical.
A similar absence of scaling had been observed for the conventional BTW model on Euclidean lattices without dissipation at boundaries \cite{grassberger}.

\begin{figure}[t]
\centering
\includegraphics[width=1.\columnwidth]{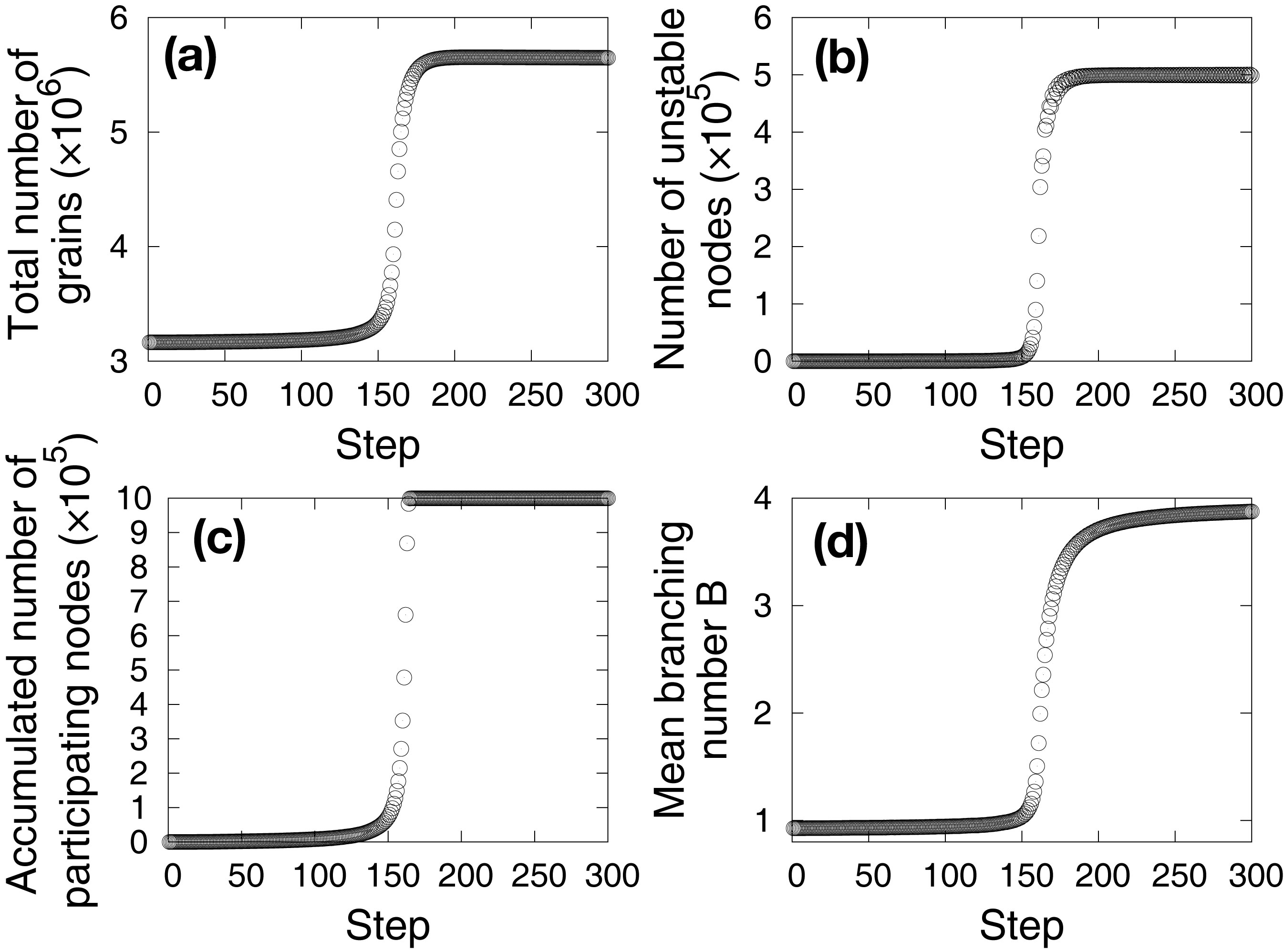}
\caption{Evolution of an infinite avalanche in the 1$\circledast$2 model. (a) The total number of grains piled in the system at a specific step of the infinite avalanche. (b) Number of unstable nodes at a specific step. (c) The accumulated number of nodes participating in the infinite avalanche up to the specific step. (d) The mean branching number $B$ at the specific step of the infinite avalanche. }
\end{figure}

\section{Summary and conclusion}
We have studied a number of variants of the Bak-Tang-Wiesenfeld sandpile model on multiplex networks. To account for the multiplexity of real-world complex systems, we introduced the multiplex sandpile model  in which the network layers are interdependent so that a node fails if it becomes unstable in any layer (called the 1$\otimes$2 model). We found that this multiplex dynamics did not alter the scaling behavior of the mean-field-type sandpile dynamics, yet the detailed pattern of the microscopic cascade dynamics at the individual node level was affected by the multiplexity. Compared to the simplex counterpart, in multiplex dynamics, higher-degree nodes such as hubs in scale-free layers are more likely to fail. This finding resonates with the recent finding of higher vulnerability of scale-free interdependent networks than Poisson-random ones under mutual percolation \cite{buldyrev10}. The robustness of scaling behaviors is further tested by considering other variants of multiplex dynamics. We could alter the scaling behavior only by breaking the conservation law, introducing excess stress due to interdependency, which drives the system supercritical. In this sense, the scaling properties of the BTW sandpile model is strongly robust on networks. In order to alter its scaling properties, one needs to alter the asymptotic form of $q_k$, which is strongly constrained by the underlying substrate network structure. Therefore, random multiplex couplings between finite number of layers would not likely induce a nontrivial scaling behavior of BTW-type sandpile dynamics despite differences in the microscopic dynamics. We note, however, that for other classes of cascade dynamics, the multiplexity can play a critical role \cite{brummitt}. 

\begin{acknowledgements}
This work was supported in part by Mid-career Researcher Program (No.~2009-0080801) and Basic Science Research Program (No.~2011-0014191) through NRF grants funded by the MEST. K-ML is supported in part by Global Ph.D. Fellowship Program (No. 2011-0007174) through NRF, MEST.
\end{acknowledgements}

\end{document}